\pdfoutput=1
\documentclass[10pt]{article}

\usepackage{color}
\usepackage{bbm}
\usepackage{epsfig}
\usepackage{amssymb,amsmath,amscd, mdwmath}
\usepackage{graphicx,cite,url}
\usepackage{algorithmic}
\usepackage{float}
\usepackage{amsfonts}
\usepackage{syntonly}
\usepackage{multirow}
\usepackage{graphicx}
\usepackage{subfig}
\usepackage{tabularx}
\usepackage{setspace}
\usepackage{stfloats}
\usepackage{amsthm}
\usepackage[labelfont=bf,labelsep=period,justification=raggedright]{caption}

\makeatletter
\renewcommand{\@biblabel}[1]{\quad#1.}
\makeatother

\date{}

\begin{document}

\begin{flushleft}
{\Large
\textbf{Exploring cooperative game mechanisms of scientific coauthorship  networks}
}
\\
Zheng Xie$^{1, \sharp }$
Jianping Li$^{1}$
Miao Li$^{2,3}$
\\
\bf{1}   College of Science,   National University of Defense Technology, Changsha,   China\\
\bf{2} School of Foreign Languages, Shanghai Jiao Tong University, Shanghai,   China \\
\bf{3} Faculty of Arts, Campus Sint-Andries, KU Leuven,    Antwerp, Belgium
\\  $^\sharp$ xiezheng81@nudt.edu.cn
 \end{flushleft}
\section*{Abstract}


Scientific coauthorship, generated by  collaborations and competitions among researchers, reflects effective organizations of human resources.
Researchers, their expected benefits  through collaborations, and their cooperative costs  constitute the elements of a game. Hence we propose a cooperative game model to explore the evolution mechanisms of scientific coauthorship networks. The model generates geometric hypergraphs, where the costs are modelled by space distances, and the benefits are expressed by node reputations, i.~e. geometric zones that depend on node position in space and time. Modelled cooperative strategies conditioned on positive benefit-minus-cost reflect the spatial reciprocity principle in collaborations, and   generate high clustering and degree assortativity,
 two typical features of coauthorship networks. Modelled reputations generate the generalized Poisson parts and
fat tails appeared in specific  distributions
of empirical data, e.~g. paper team size distribution. The combined effect of modelled  costs and reputations reproduces the transitions emerged in degree distribution, in the correlation between degree and local clustering coefficient, etc. The model provides an example of how individual strategies induce network complexity, as well as  an application of
   game  theory to social affiliation networks.




\section*{Introduction}


Collaborations  between researchers  contribute not only to the breakthrough achievement  unattainable by individuals~\cite{Borner3,Jones},
 but also to the transmission and fusion of knowledge, and hence  they incubate several interdisciplines~\cite{Adams,Shrum,Uzzi, Wuchty}.
 Coauthorship in scientific papers, as a valid proxy of collaborations,
 can be expressed graphically (termed  as coauthorship network), where nodes and edges represent authors
 and coauthorship respectively. Studies of large-scale coauthorship networks
  provide a bird-eye view of collaboration patterns in
diverse fields, and have become an important topic of social sciences~\cite{Glanzel1,Glanzel2,Sarigol,Mali,Jia}.

Empirical  coauthorship networks have  specific  common local (degree assortativity, high clustering) and global (fat-tail, small-world)   features~\cite{Newman1,Newman2,Newman3,Newman0,Newman4,XieLL}.
 Some important models have been proposed to reproduce those properties, such as modeling   fat-tail   through   preferential attachment or cumulative advantage~\cite{Barab,Moody,Perc,Tomassini,Wagner,Santos}, modeling   degree assortativity   by connecting two    non-connected nodes that have  similar   degrees~\cite{Catanzaro}. Except for preferential attachment,
the inhomogeneity of node influences is an alternative explanation for fat-tail: Nodes with wider influences are likely to gain more connections~\cite{Krioukov1}.
The idea has been applied to model coauthorship
  networks in a geometric way: Node influences are modelled by attaching   specific   geometric zones to nodes~\cite{Xie1,Xie3}.


To find the essence from the above features, we
face a basic  question\cite{sci125}: ``How did
cooperative
behavior evolve?"  Five typical  mechanisms of cooperative evolution\cite{Nowak} all hold for coauthorship:
Coauthoring  frequently occurs between students and their tutors (Kin   selection);
Cooperation  helps to achieve breakthroughs that are unattainable
by individual (Direct reciprocity);
Coauthoring someone could establish  a good reputation (Indirect reciprocity);
  Spatial structures or social networks make some researchers interact
more often than others and obtain more collaborators (Network reciprocity);
   A successful research team  is attractive for collaborators (Group selection).
  To quantify collaborations and predict
behavioral outcomes, a
modelling approach termed as game theory is developed to find rational strategies.
Then do there exist    inherent game rules behind the   complexity of   coauthorship  networks?


We  try to find a solution for the above question  through simulation.
A cooperative game consists of two elements: a set of players  and a characteristic function specifying
the value (i.~e. benefit-minus-cost) created by subsets of players in the game.
Scientific cooperation has those elements.
The diversity of researchers' learning programs of  leads to their individual research interests.
Cooperation costs could be considered as    investments of time and effort to
complete  a study by crossing
the distance between research interests\cite{Hoekmana}.
The reputation in academic society  could be regarded as the expected benefit of cooperation: Coauthoring with a famous researcher
contributes to achieve academic success.





In the model,
   the set of interests   is abstracted as a circle, and players are   located on the circle.    Cooperative costs
are geometrized as  angular distances, and  the reputation benefit of a player is valued as a power function of player generation time. Modelled cooperative strategies conditioned on positive benefit-minus-cost imitate the spatial reciprocity principle in collaborations\cite{Hauert}, and   yield high clustering and degree assortativity. The designed form of reputations, together with the strategies,
 yield  the features (hook heads, fat tails\cite{Milojevic}) of specific distributions of empirical data, such  as degree distribution, the distribution of paper team sizes, etc. Moreover, the combined effect of spatial reciprocity and the diversity of reputations reproduce the transition phenomena in degree distribution, in the correlation between degree and local clustering coefficient, etc. The good model-data fitting shows the reasonability of the designed game mechanisms.

This paper is organized as follows: The model and   data are described in Sections 2 and 3 respectively;  Cooperation cost, reputation benefit and
 the relationship between them are discussed
   in Sections 4 and 5 respectively; The conclusion is drawn in Section 6.

\section*{The model}


Hypergraph is a generalization of   graph,  in which an edge (termed as hyperedge) can join any number of nodes.
 Coauthorship relationship can be expressed by a  hypergraph, where nodes represent authors, and the author group  of a paper (called it a ``paper team") forms a hyperedge.
 A number of models have been proposed for generating hypergraphs in specific random ways,
and some of them have been used for
 modelling coauthorship networks\cite{Borner,Milojevic,Xie7}.
Meanwhile, there has been
an amount of  previous work on the
structures of    specific  random hypergraphs, such as clustering, the emergence of a giant component, and so on\cite{GoldschmidtC,Estrada2,Darling,Xie6}.

\begin{figure}\centering
\includegraphics[height=2.4  in,width=3.8   in,angle=0]{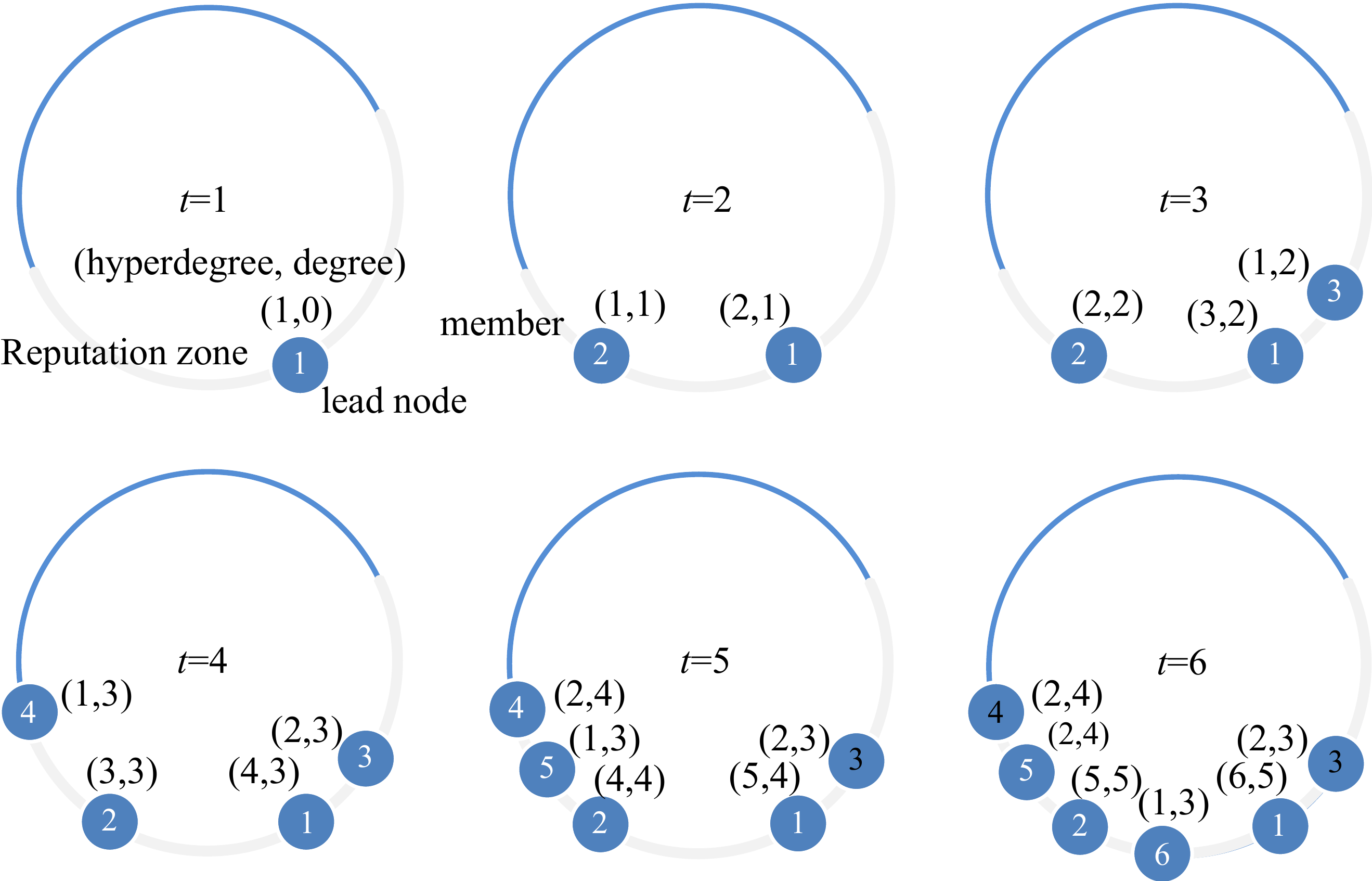}
\caption{{\bf An illustration of  an evolutionary   research team. }
 At each time, a new player joins the research team, and the group of the players with white   numbers forms a hyperedge.
 It illustrates  an usual scene: A new coming player $6$  wants to complete a work
and write it as a paper. Suppose completing the paper  needs four players. Then player $6$ would ask the leader
$1$ for help, and the leader would suggest   team members $2$ and $3$, who have most similar
interest to player $6$, to cooperate.
 } \label{fig0}      
\end{figure}

  We provide   a geometric hypergraph model, where   the set of research   interests  is abstracted as a  circle  $S^1$,
and  researchers are expressed as nodes  located      on  the circle~(Fig.~\ref{fig0}).
The nodes are generated in batches from $1$ to $T\in \mathbb{Z}^+$, hence they can be identified by   spatio-temporal coordinates.
 Some nodes are randomly selected as  ``lead nodes"  to attach specific  arcs   that  imitate    their     ``reputations".
 The nodes covered by a lead node's   arc  constitute
a  ``research team".
The paper teams are modelled by
hyperedges, which are generated by following cooperative game mechanism.

A cooperative game consists of two elements:  a set of players $N$  and  a
characteristic function specifying the value (benefit-minus-cost) created by  subsets of
players in the game.
The characteristic function is a function    $v$ mapping each subset $S$ of $N$ to  the  value   $v(S)$ it  creates.
Regard   nodes   as
 players  $N = \{i_1,...,i_n\}$. Think of player $i_l$ as a lead node with   players  $i_{l_1},...,i_{l_{m}}$ as its research team members,
and
player $i_c$ as a  candidate  attempting  to cooperate with $i_l$ and specific members (e.~g. $i_{l_1},...,i_{l_s}$).
    Assume that the cooperation  cost   is $d(i_l,i_{l_1},...,i_{l_s},i_c)$, and suppose that those players will  receive a  benefit valued by $i_l$'s reputation
$r(i_l)$. We can define an intuitive characteristic function $v$   for this game  as follows:
\begin{equation} v(S) =\left\{\begin{aligned}\label{eq1}
      r(i_l)-d(i_l,i_{l_1},...,i_{l_s},i_c),   ~~~{S= \{i_l,i_{l_1},...,i_{l_s},i_c\};}  \\
      0, ~~~~~~~~~~~~~~~~~~~~~~~~~ ~ ~~~~~~~~  { S\neq\{i_l,i_{l_1},...,i_{l_s},i_c\}.}
\end{aligned}\right.
\end{equation}
Under the definition (\ref{eq1}), if $v(\{i_l,i_{l_1},...,i_{l_s},i_c\})>0$,   those players  will collaborate. 



The empirical distributions of  paper team sizes emerge  a hook head and a fat tail,
 which means the sizes of substantial papers  are around their average, and a few papers have a  significantly  large size.
 In reality,   researchers in a small research team
 are more likely to write papers
together. Members of a large research team rarely coauthor a paper all together, but rather  with  a   fraction of members.
Treating paper team size as a random variable $x$,
 we design a mechanism to simulate the distribution of
  $x$.
Give   the upper bound of small research team  $\mu>0$, and the lower bound   of large research team  $\nu>0$.
 Denote the expected value of paper team size and the size of
 corresponding
 research team     to be $\eta$ and $\lambda$ respectively.
Let  $\eta    = \lambda  $, if   $\lambda\leq\mu$;
 Let  $\eta  =\mu$, if   $\mu<\lambda \leq\nu $;
Draw $\eta $  from a power law distribution with  an exponent $\gamma$ and domain $[\mu,\lambda]$, if   $\lambda>\nu $.
Then  draw
  $x$   from a Poisson distribution with   expected value $\eta$.
Note that in the description of the above  game,
  $x=s+2$   and $\lambda=m+2$.



Cooperation  costs   could be considered as   investments of time and effort to
complete a study by crossing
the distance between the research interest of  the leader and that of the candidate, etc.
Denote the spatio-temporal coordinates of player $i\in N$ by $(\theta_i,t_i)$, and write the player as $i(\theta_i,t_i)$.
We abstractly     geometrize the cost $d(i_l,i_{l_1}...,i_{l_s},i_c)=\pi-|\pi-|\theta_{i_l}-\theta_{i_c}||$, namely
the angular distance between $i_l(\theta_{i_l},t_{i_l})$ and $i_c (\theta_{i_c},t_{i_c})$.

We now show how to value reputation.
Considering the inefficient  information    of new players,
we simply assume each lead node has the same attraction to new players  and so   value the reputation of  a lead node $i (\theta_{i},t_{i})$ as
     $r( i)\propto 1/t_i $. Hence  the
 expected number of
 $i$'s collaborators    $k_i(t)=\alpha (T-t_i)/t_i\approx \alpha T/t_i$ at time $T\gg t_i$, where $\alpha>0$. Those   yield   $P(k_i(T)<k) =  P (t_i >  \alpha  T/k) $.
  The probability
density of a lead node  generated at $t_i$ is $1/T$. Hence
 $ P (t_i >  \alpha t/k) =1-P (t_i \leqslant \alpha T/k)=1-\alpha  / k  $. Then
the tail of degree distribution
$P(k)=\partial P(k_i(T) < k)/\partial k\propto 1/k^2$.
We can obtain the general case
$P (k) \propto 1/k^{1+1/\beta}$ for large enough  $k$   by
valuing    the reputation
 $r(i)={\alpha}t^{-\beta}_it^{\beta-1}$, where $\beta\in(0,1]$.
The strict mathematical deduction of the degree distribution tail
 needs  averaging on Poisson distribution, which is inspired by some of the same general ideas
as explored in Ref.~\cite{Krioukov1}.

We  next  show how to generate a paper team, namely    cooperation rule.
  Empirical collaboration behaviors have specific certainty  (due to
kin selection, network selection, etc.), as well as    uncertainty.
Consider an usual  scene,
a researcher  $i$   of leader $j$'s  research team wants to complete a  work and   write it as a paper, which needs $x\geq 2$ researchers to work together.
Then $i$ would ask  the leader   $j$ for help, and
  $j$ would  suggest $(\min(x, |R_j|)-2)  $  members
of his research team $R_j$, who have most similar interest to $i$, to cooperate with $i$.
Such   behavior can be viewed as kin selection, and  is featured in certainty.
When finishing the  work is beyond the ability of  the   team $R_j$, the researcher $i$ would ask for external helpers.
Uncertainty exists in
this selection behavior, which inspires the design of randomly choosing  $(x-\min(x, |R_j|)  $  players outside of $   R_j $
 to cooperate.
 The uncertainty shorts the  average shortest path length   of  modelled networks.
   Note that  a researcher   could belong to several research teams, hence above scene would happen in each team.


   Based on   above set-up, we
build the hypergraph model  as follows:
 \begin{description}
\item[1.] Reputation assignment
\item For time $t=1,...,T\in \mathbb{Z}^+$ do:
\subitem
  Sprinkle   nodes $N_t$   as new players   uniformly and randomly on $S^1$.
  Select subset  $N^l_t$ from $N_t$ randomly  as lead nodes, and value the reputation of $j (\theta_j,t_j)\in N^l_t$
     as
    $r(j)={\alpha}t^{-\beta}_jt^{\beta-1}$.
\end{description}

\begin{description}
\item[2.]  Cooperation rule
\item For time $t=1,...,T$ do:
\subitem  For each new node $i(\theta_i,t_i)\in N_t$, select a lead node set $M^l_i$ for which   $\forall j (\theta_j,t_j)\in M^l_i$ satisfies $r(j)>\pi-|\pi-|\theta_{i}-\theta_{j}||$ and $t_j<t_i$.
For  each  $j\in M^l_i$, add $i$ to $j$'s  research team $R_j$,
and  generate a hyperedge at probability $p$ by grouping   $i$, $j$, $(\min(x,|R_j|)-2)$  players of $R_j$   nearest to $i$, and
$(x-\min(x,|R_j|))$     players  $\not\in R_j$ randomly,   where   $x$ is the random variable  above defined.
\end{description}

%

The player set   of the model $N=\bigcup^T_{t=1} N_t$,  and the number of players  $n=|N|=\sum^T_{t=1} |N_t|$. Here we let  $|N_t|$ and   $|N^l_t|$ be
constants over  $t$.
Compare with the model in Ref.~\cite{Xie6}, the new model reduces the number of parameters. Moreover,  the new model
has the  ability to reproduce the empirical feature  of the distribution  of hyperdegrees,   and  that of paper team sizes.
A node's    hyperdegree    is the number of hyperedges that contain  the node.

\section*{The data}
To test the fitting  ability   of the proposed model, we analyze two   empirical coauthorship networks~(Table~\ref{tab1}). Dataset  PNAS  is composed of   52,803     papers published in   {\it Proceedings of the National Academy of
Sciences} during 1999--2013.
Dataset PRE comprises 24,079 papers published in {\it Physical Review E}    during  2007-2016.
 Note that  43,304 papers of the first dataset belong to biological sciences,
and the second dataset comes from  physical sciences.
The different collaboration level (reflected by the average number of authors per paper) of the two datasets (PNAS 6.028, PRE 3.102)
helps to test  the flexibility
 of the model.

In the process of  extracting networks from those metadata, authors are identified by their names on their  papers.
For example, the author named  ``Carlo M. Croce" on his paper is represented  by the name. We mainly focus on the
distribution of degree and that of hyperdegree  as well as some properties based on degrees.
   From the  analysis of Ref.~\cite{Kim1}, we find  that
  identifying authors by their name  on papers holds the
degree distribution feature
of   ground truth  data, which
    partially verifies  the reliability of the   empirical networks used here.

  Using surname and the initial of the first given
name   generates a lot of merging errors of  name disambiguation\cite{Milojevic3}. Hence we compute the proportion
of those authors,
and that of those   authors further conditioned on publishing more than one paper.  Meanwhile, Chinese names were also found to
account for the repetition of names\cite{Kim1}. We count the
proportion of names with a given name less than six characters and a surname
among major 100 Chinese surnames.   The small proportions of such   authors
and
those of such authors publishing more than one paper~(Table~\ref{tab1})
  limit
  the impact of name repetition, especially for dataset PNAS.

\begin{table*}[!ht] \centering \caption{{\bf Specific   statistical   indexes of  the empirical data.} }
\begin{tabular}{l r r r r r r r r r r} \hline
Data&     $a$ & $b$ & $c$ & $d$\\ \hline
  PNAS  &  2.62\%&  1.08\%  & 2.90\% &0.27\% \\
PRE  &3.85\% &1.58\%  &19.2\%&6.45\%    \\
\hline
 \end{tabular}
  \begin{flushleft}   Indexes $a$ and $b$ are
  the percent of authors who have a surname among major 100    Chinese surnames and only one given name shorter than six characters,
 and the percent of those authors further conditioned  on  publishing  more than one papers respectively.
Indexes $c$ and $d$  are the percent of  authors who use surname and the initial of the first given
name,
 and the percent of those authors further conditioned   on  publishing  more than one papers respectively.
\end{flushleft}
\label{tab1}
\end{table*}

To reproduce specific  features  of the empirical  data, we
choose proper parameters~(Table~\ref{tab2}) to
generate
 two  hypergraphs, and extract  simple
graphs from them (where   edges are formed between every two nodes in each hyperedge,
isolated nodes are ignored, and   multiple edges are viewed as one).
Since the model is stochastic, we generate 20 networks with the same parameters, and   compare their statistical   indicators in Table~\ref{tab3}. The finding is that  the model is robust on those  indicators~(Table \ref{tab4}).

%



  \begin{table*}[!ht] \centering \caption{{\bf The parameters of Synthetic-1, -2.} }
\begin{tabular}{l llllllll  } \hline
  $T=5,000, 6,000$ & $N_1=100, 15$&$N_2=5,   5$& $p=0.25, 0.4$ & \\
   $\alpha=0.13, 0.2$& $\beta=0.52, 0.55$ & $\gamma=4.2, 4.2$& $\mu=6, 2$& $\nu=42, 6$& \\
\hline
 \end{tabular}
  \begin{flushleft}
The parameters in the first row control   network size, and those in the second row control a range of distributions, such as degree
distribution, hyperdegree distribution, etc.
\end{flushleft}
\label{tab2}
\end{table*}



 \begin{table*}[!ht] \centering \caption{{\bf Specific   statistical   indexes of  the analyzed networks.} }
\begin{tabular}{l r r r r r r r r r} \hline
Network&NN&NE   & GCC & AC &AP & MO & PG     \\ \hline
  PNAS &201,748&1,225,176 &0.881 &0.230 & 6.422 &0.884 &  0.868   \\
Synthetic-1   &128,749 & 694,769& 0.864& 0.229  &11.35&0.987&0.648   \\
PRE & 37,528 &90,711&  0.838&  0.394 &6.060  &0.950&  0.583 \\
Synthetic-2   &29,397& 62,834&  0.829 &  0.174 &12.91 &0.983 &0.505  \\
\hline
 \end{tabular}
  \begin{flushleft} The indexes are    the numbers of nodes (NN) and edges (NE),  global  clustering coefficient (GCC),  assortativity coefficient   (AC),   average shortest path length (AP),  modularity (MO), and the node proportion of the giant component~(PG).
  The values of AP  of   the first two networks are calculated by sampling 300,000 pairs of nodes.
\end{flushleft}
\label{tab3}
\end{table*}

\begin{table*}[!ht] \centering \caption{{\bf The  means  and   standard deviations (SDs) of specific   indexes.} }
  \begin{tabular}{l r r r rr r r r r r r r} \hline
Synthetic-1&NN&NE    & AC &GCC & PG     &MO & AP  \\ \hline
Mean& 1.29E$+05$ &	6.23E$+04$&	2.45E$-01$&8.64E$-01$&	6.53E$-01$&	9.87E$-01$&	1.12E$+01$  \\
 SD &7.66E$+02$&	1.14E$+04$	&3.45E$-02$	&7.08E$-04$&	8.03E$-03$	&4.97E$-04$	&1.47E$-01$\\\hline
Synthetic-2&  \\ \hline
Mean& 2.93E$+04$ &	6.23E$+04$&	1.35E$-01$&	8.30E$-01$&	5.06E$-01$&	9.82E$-01$&	1.22E$+01$  \\
 SD &1.48E$+02$&	5.89E$+02$	&2.93E$-02$	&6.13E$-04$&	6.48E$-03$	&6.18E$-04$	&5.33E$-01$
 \\
\hline
 \end{tabular}
   \begin{flushleft} The meanings of   headers are shown in the notes of Table~\ref{tab3}.
\end{flushleft}
\label{tab4}
\end{table*}

 \section*{Cooperation cost  and   reputation benefit}
 \label{sec4}

%
%





Based on
the cost and the benefit of collaborations, we explain    the  distribution feature   of paper team sizes.
The benefit of joining a paper team is limited. The   law of diminishing marginal utility holds in academic society.
The allocation of academic achievements is often according to author order.
   Hence only the researchers with positive
  benefit-minus-cost would   join the paper team. Assume  the number  of those researchers is $n_r$.
Meanwhile,  the joining behaviour   has certain degrees of randomness. Let the  joining probability    be $p$.
Then  the paper team size  will follow  a binomial distribution, and so    a Poisson distribution  with   expected value
$n_r{p}$  approximately~(Poisson limit theorem).
  Due to the   law of diminishing marginal utility,  the sizes
  of those papers would
   follow  a    generalized Poisson distribution, because this distribution  describes situations where
         the occurrence probability of  an event involves
memory~\cite{Consul}.


    Some important works require many researchers (even  from different research teams) work  together, which   would  bring about huge economic and social benefits.
     The papers of those works would have many authors,      and sometimes show their appearances  in specific famous journals, e.~g. a paper in {\it Nature} has 2,832 authors~(see Fig. 3 in Ref.\cite{Xie7}).  In fact, signing on a paper of a famous journal will also bring about  a huge    benefit.
 The existence of those   papers   leads to     fat
tails emerging paper team size distributions.

\begin{figure}\centering
\includegraphics[height=3.2  in,width=6.   in,angle=0]{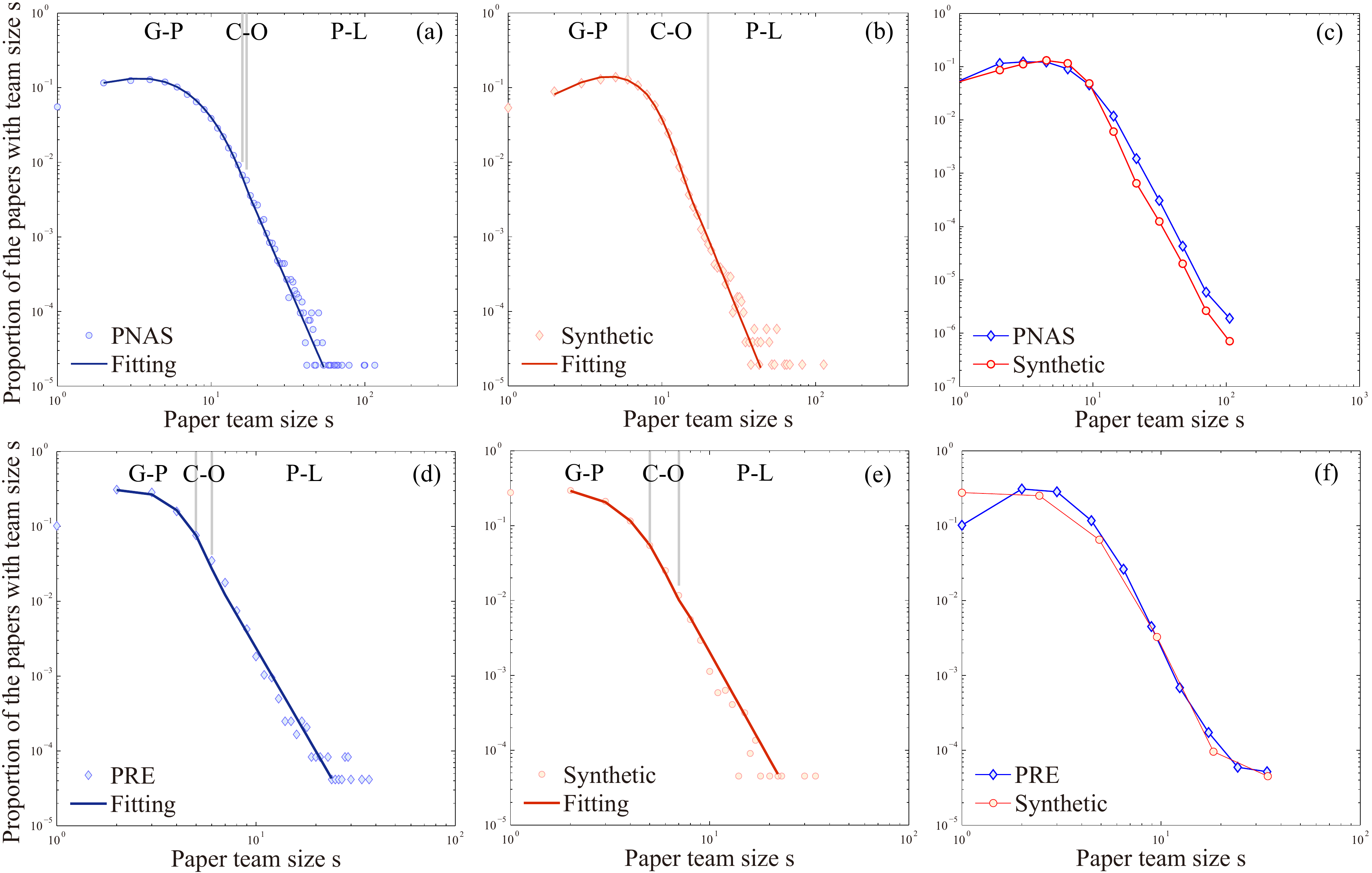}
\caption{{\bf The distributions   of sizes per paper team.   }
In the panels of the first two columns, the regions ``G-P",  ``C-O", ``P-L" stand  for generalized Poisson, cross-over  and power-law respectively. The parameters and goodness of fittings are listed in Table~\ref{tab7}.
 } \label{fig1}      
\end{figure}

In brief,
above analysis makes us think that
 benefit-minus-cost and the randomness of joining behaviour make a paper team size
 follow
 a  generalized Poisson distribution,
and  huge expected benefits lead to fat-tail.
There exists a cross-over between the two limits~(Fig.~\ref{fig1}).
 The fitting function of the distribution, including  following discussed distribution  of hyperdegree  and that of degree,
is a combination of  a generalized Poisson distribution and a power-law function~(Table~\ref{tab7}).
 We perform a two-sample Kolmogorov-Smirnov (KS) test
to compare the distributions of   two data vectors:   indexes (e. g. paper team sizes), the samples drawn from the corresponding
fitting distribution. The null hypothesis is that the two data vectors are from
the same distribution. The $p$-value of each fitting shows the test cannot reject
the null hypothesis at the 5\% significance level.

\begin{table*}[!ht] \centering \caption{{\bf  The parameters and goodness   of fitting functions for degree distribution, hyperdegree distribution, and the distribution  of paper team sizes, from the top down.} }
\begin{tabular}{l rr r r rr r r r r r r r} \hline
    &$a$ &$ b$   & $c$ &$d$  &$s$ & $B$   & $E$      &$p$-value  \\ \hline
 &    4.751 &   0.434   &  110.8 & 2.987 &  0.845&  10 &25 &    0.256\\
PNAS &   0.085  & 0.367 & 2.368 &2.941&11.99& 3 &7 &  0.128  \\
   &    3.438&0.382  & 3,094 &4.755&   1.095& 16 &17 & 0.174  \\
\hline
 &   5.082   &    0.380 &   708.4 & 3.580  &  0.898&  13 &32 & 0.074\\
Synthetic-1 & 0.664  &0.365  &  16.15 &3.694 &1.999  & 4 &9 &  0.971  \\
 &  4.502  & 0.208    &3,621 &5.059&  1.003 &  6 &20 & 0.103  \\
\hline
 &     2.776   &   0.156 &   8.910 & 2.758  &  0.898&  5 &6 & 0.118\\
PRE &    0.038 &0.399  &   4.260 &3.079  & 25.56   &3 &10& 0.875   \\
  &    2.507   & 0.000   &82.66& 4.587 & 1.249 &  6 &7 &0.133  \\
 \hline
 &   2.352  &  0.2367 &  46.65 & 3.460  & 0.970&  5 &11 & 0.052\\
 Synthetic-2 &   0.375 &0.272 &  1.786 &2.991 & 3.077  &5 &6& 0.673  \\
&  2.007    &  0.065    & 116.1& 4.756 &1.146  &  5 &7 &0.887  \\
\hline
 \end{tabular}
  \begin{flushleft}
  The domains of  generalized Poisson  $f_1(x)= {a (a+bx)^{x-1}}{ \mathrm{e}^{-a-bx }/{ x!}} $  ,  cross-over and power-law   $f_2(x)=cx^{-d}$  are $[\min(x),E]$,  $[B, E]$ and  $[B,\max(x)]$ respectively.
  The fitting function defined on $[\min(x),\max(x)]$ is $f(x)=q(x)      s f_1(x) +(1-q(x))f_2(x)$, where
  $ q(x)=\mathrm{e}^{ - (x -B)/(E-x ) }$.    The fitting processes are:
  Calculate parameters of $sf_1(x)$   and $f_2(x)$  by regressing  the head and tail of  empirical  distribution  respectively;      Find  $B$ and $E$ through exhaustion    to  make $f(x)$ pass   KS   test ($p$-value$>0.05$).  The $p$-value   measures the  goodness-of-fit.
   \end{flushleft}
\label{tab7}
\end{table*}


In the model, with a proper upper bound parameter $\mu$ (around average number of authors per
paper of corresponding empirical data),
the
model  can reproduce
 the generalized Poisson part of  the distribution  of paper team sizes,
because most  of  modelled paper team sizes are drawn from Poisson distribution with an expected value around $\mu$.
  Meanwhile,   with a proper lower bound parameter $\nu$,
the mechanism
 can generate a few    significantly large paper team sizes, and so  the fat tails of the modelled paper team size distributions.
We choose $\nu$ through  iteration from  the starting point   of the power-law part in
   the corresponding empirical distribution of paper team sizes ($E$ in Table~\ref{tab7}) until the modelled networks  have  the similar  feature  of  the
   empirical   distribution of degrees
    and that of paper team sizes.

%


Now
we turn  to  explain the distribution feature of   degrees.
Substantial authors  publish only one paper  (PNAS: 64.8\%, PRE: 63.9\%),
and most of paper team sizes draw from a generated Poisson distribution (PNAS: 99.9\%,  PRE: 99.9\%).
Those  lead  the generalized Poisson parts of degree distributions.
Note that the boundaries of generated Poisson parts of   paper team size distributions  are 41 and 20 for PNAS and PRE respectively, which
are detected by the boundary point  detection algorithm for  probability density functions  in Ref.\cite{Xie6}  (listed in Appendix).

\begin{figure}\centering
\includegraphics[height=6.4  in,width=6.   in,angle=0]{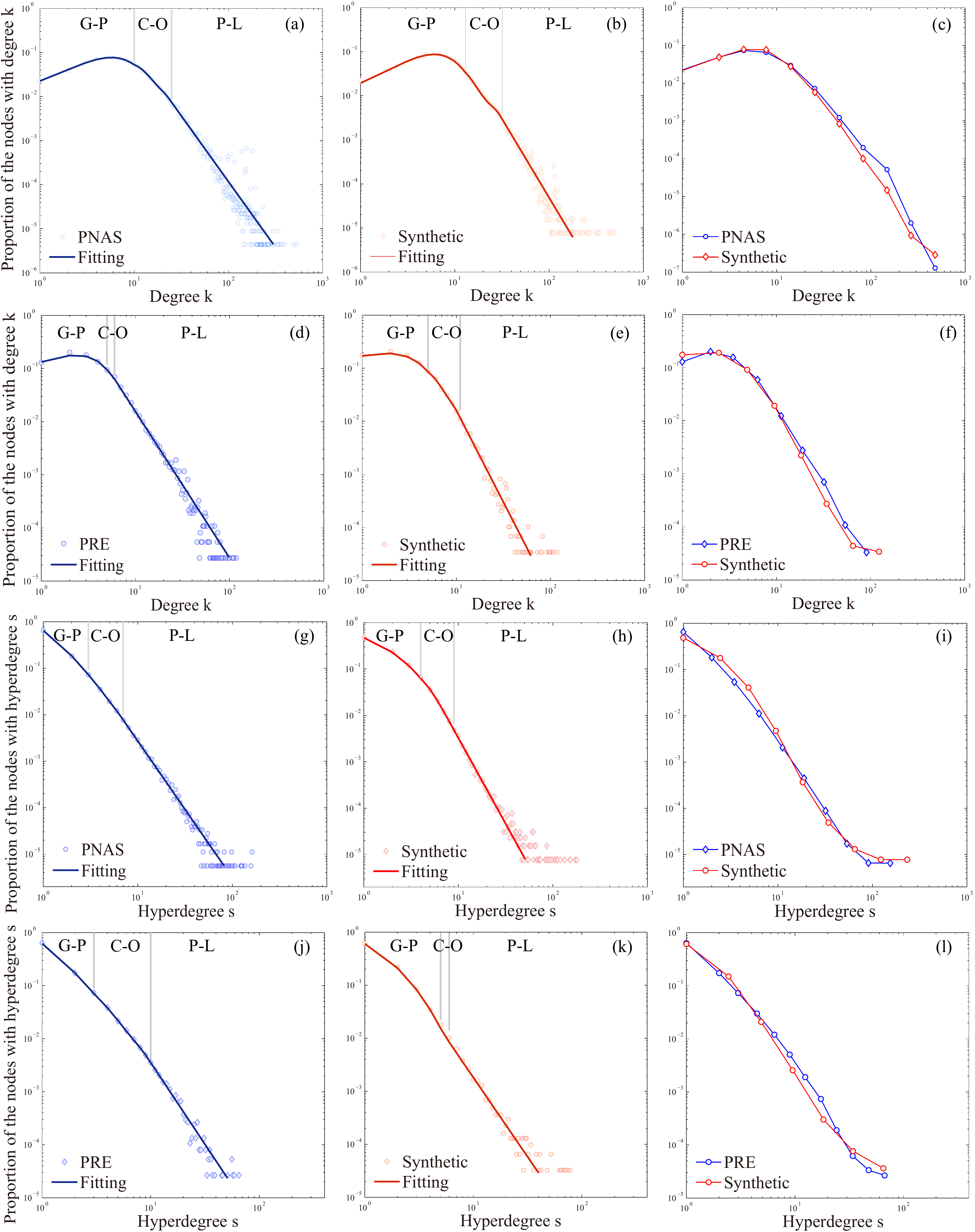}
  \caption{ {\bf The empirical  and synthetic   distributions of  collaborators/papers  per author.} The regions ``G-P",  ``C-O", ``P-L" stand  for generalized Poisson, cross-over  and power-law respectively. The parameters and goodness of fittings are listed in Table~\ref{tab7}. }
 \label{fig2}      
\end{figure}

With the growing of their papers, a few authors   experience  the cumulative process of collaborators over time, whose
   reputations also increase. As  empirical data show,
 it is an accelerative process, which is often explained by    cumulative advantage.
The  process
reflects as the transition from   a generated Poisson   to  a   power-law~(Fig.~\ref{fig2}).
  Above explanation  can also be used to explain the similar feature of hyperdegree distributions.
Note that  the nodes of  large   paper teams   also have a large degree,
which reflects  as the outliers in the tails of degree distributions.

%



In the model, we can
 choose suitable  parameters $\alpha$,  $|N_t|$, $|N^l_t|$ and $p$ to make  the hyperdegrees of substantial players be one~(Synthetic-1:  48.5\%,
Synthetic-2:  61.5\%).
Meanwhile, the substantial   modelled paper team sizes follow  a generalized   Poisson distribution~(Synthetic-1: 99.9\%, Synthetic-2: 100\%). Those yield the generalized   Poisson part of   modelled degree distributions. The boundary
of generalized   Poisson part
is   34 for Synthetic-1 and   23 for Synthetic-2.
 The mechanism of generating hyperedges makes only  early  lead nodes and specific players  close to them  can  experience  the cumulative process of connecting new players.
The cumulative process  generates the fat tails   of   modelled degree distributions, as well as those of   modelled  hyperdegree distributions~(Fig.~\ref{fig2}).
The cumulative  speed and so the power law exponent  can be tuned by   parameter $\beta$.

\section*{Spatial reciprocity   and network reputation}




Cooperation needs to be based on acquaintanceship. Hence there is an acquaintanceship network under each coauthorship network. Geographic contexts (such as organization, institution, etc.) contribute to emerging clustering structure in acquaintanceship network, namely ``the friend of my friend is also my friend"~\cite{Newman4}. The Internet extends the scope of acquaintanceship, which crosses spatial barriers even national boundaries. Therefore, the factor of clustering changes from geography to interest, namely ``birds of a feather flock together"\cite{McPherson}.


Cooperation costs make cooperators should have similar research interests,
namely collaborations exist  in     researcher clusters formed by similar interests.
 Hence the spatial reciprocity principle in cooperative game theory\cite{Hauert} needs to be modified by interest in the situation of academic cooperation.
In a network perspective,  the extent of spatial reciprocity can be reflected by
  local  clustering coefficient and
the degree difference between  a node
 and its neighbors.





\begin{figure}\centering
\includegraphics[height=6.4  in,width=6.   in,angle=0]{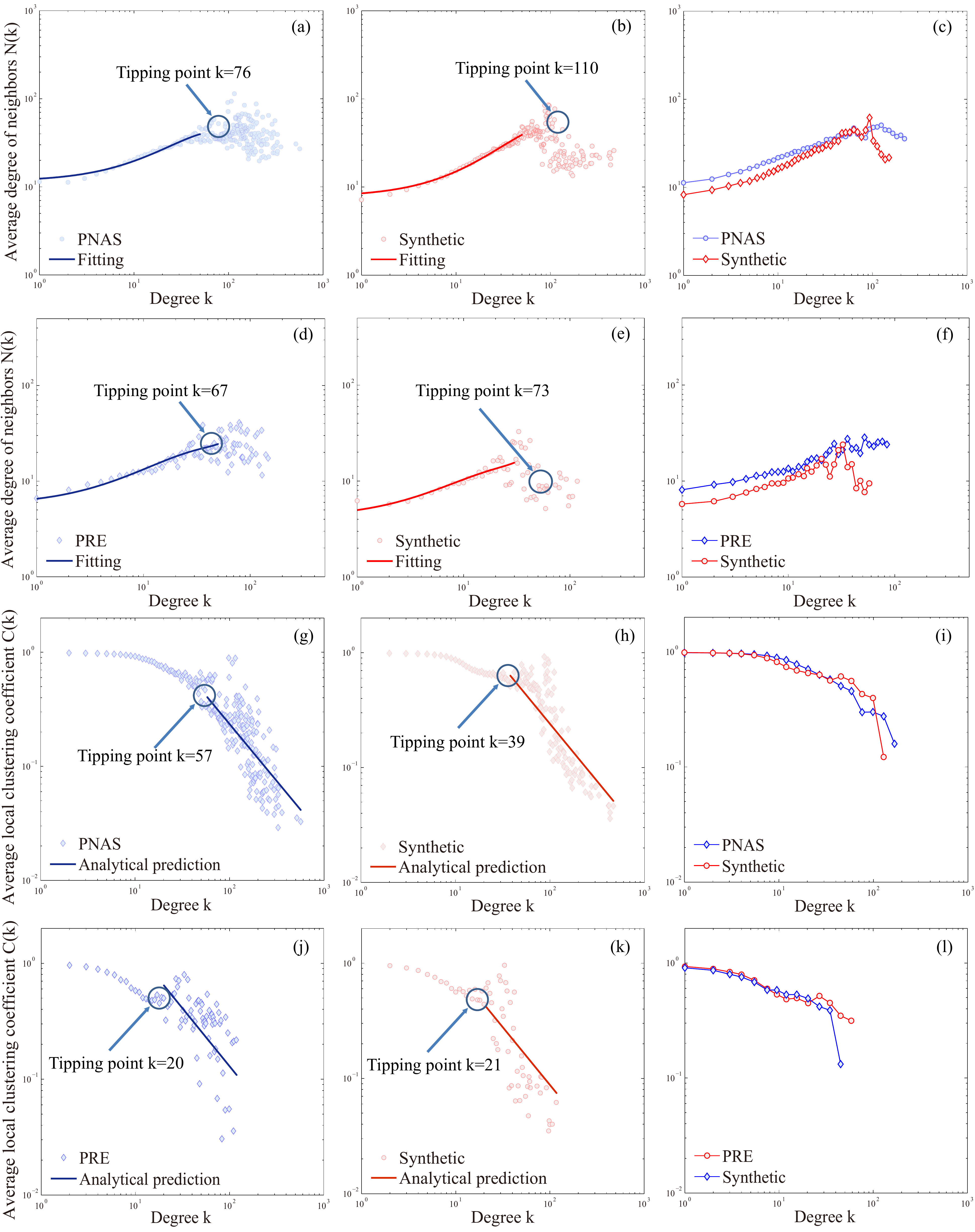}
\caption{  { \bf The    relation between degree and  average degree of neighbors,
and that between local clustering coefficient and  degree.
} The panels  show    $k$-degree nodes' average degree of  their neighbors and their
the average  local clustering coefficient  respectively.
 } \label{fig4}      
\end{figure}

Now we discuss the relationship between
  spatial reciprocity   and    network reputation.
In the view  of nodes,
 network reputation  can be reflected
by degree.
Hence the relationship
can be reflected  by two functions of degree, namely the  average local  clustering coefficient of $k$-degree nodes $C(k)$,
and  the average degree of $k$-degree nodes'  neighbors  $N(k)$.
There is a transition  in   each of the functions~(Fig.~\ref{fig4}).
The tipping points of  $C(k)$ and $N(k)$
are detected by the boundary point  detection algorithm for general functions in Ref.\cite{Xie6}  (listed in Appendix).
Inputs of the algorithm are $C(k)$/$N(k)$, $g(\cdot)=\log(\cdot)$  and $h(s)=a_1 \mathrm{e}^{-((s-a_2)/a_3)^2}$/$h(s)= a_1 s^3 + a_2 s^2 + a_3 s  + a_4$
($s$, $a_i\in  {\mathbb{R}}$, $i=1,...,4$).
  Using  those inputs is based on   the observation of $C(k)$ and $N(k)$.

Coauthorship networks are found to have
two features: high   clustering  (a high probability of a node’s two neighbors connecting) and   degree assortativity (a positive correlation coefficient between two random variables: a node's degree and the average degree of the node's neighbors), which are measured by GCC and AC  in Table~\ref{tab3} respectively.
To understand the essence  of     high   clustering, degree assortativity,  as well as the transitions in $C(k)$ and $N(k)$, we
 analyze the feature of the basic context of collaborations, i.~e. research teams.  Given the cost and   benefit of joining a research team, only the researchers with positive benefit-minus-cost would join the team with a probability (that would be affected by previous members due to  gossips, etc.). With an argument similar to the one used in the distribution of paper team sizes, we can assume the research team sizes follow a generalized Poisson distribution. A few   research teams with a huge reputation would attract substantial collaborators and become significantly large ones.

 Based on the above analysis, we can think that small degree authors comprises two parts: one is composed of the authors of small research teams, and other one comprises the unproductive authors belonging to small paper teams and to large research teams. Researchers  in the small research team probably write a paper together, which causes they have a high local clustering coefficient, and a slight degree  difference between them  and their neighbors. Many authors in large research teams only  write one paper, and the paper team only contains a few leaders. Hence those authors would have a relatively high local clustering coefficient, and a relatively small difference between their degree and the average degree of their neighbors. From above analysis, we can infer   small degree authors contribute to degree assortativity and high clustering of coauthorship networks, which fits the empirical data~(Figs.~\ref{fig4}).

The collaborators of some productive authors    may not   coauthor,
and some productive authors often have many collaborators.
 The degree difference    emerges between    those authors and  their neighbors, on average. Hence we can infer those
  large degree authors negatively contribute to degree assortativity and high clustering.
The inference  fits the empirical data:    the  tails of
  $C(k)$ and $N(k)$ of each empirical network  emerge  a  different trend from the heads~(Figs.~\ref{fig4}).
Note that the
authors of large paper teams also have a large degree, but
contribute to   degree assortativity and high clustering.
The existence of those authors  causes    the scattered  points of the tails of
 $C(k)$ and $N(k)$.

 The model can generate  research teams with a size distribution  as  above inferred.
Due to    the power function of reputation,
 the expected size $\mathbb{E}\lambda(T)$ of  a research team of lead $i$ is proportional to $\sum^T_{s=t_i+1}{\alpha}t^{-\beta}_is^{\beta-1}\approx{\alpha}(T/t_i)^{\beta} $
  for $T\gg t_i$.
This   yields   $P(\mathbb{E}\lambda(T)<\kappa) =  P (t_i >  (\alpha/\kappa)^{1/\beta}  T ) $.
 With
an argument similar for the reasonability of reputation  function,    we can obtain
the tail of the  distribution of   modelled research team sizes
$P(\kappa)=\partial P(\mathbb{E}\lambda_i(T) < \kappa)/\partial \kappa\propto 1/\kappa^{1+1/\beta}$.
 When    $T\not\gg t_i$, the research team size $\lambda$ is drawn from a Poisson distribution with an expected value proper to ${\alpha}(T/t_i)^{\beta}$ due to
  the Poisson point process of generating nodes. Hence the small modelled research team sizes are drawn from
  a range of Poisson distributions with expected values taking from a power function.
With proper parameters, those can be used as basis to fit    a given generalized Poisson distribution.

 Most   modelled  hyperedges are generated by grouping a small fraction (around $\mu$) of nodes  close in space,  which
expresses the spatial reciprocity principle. Moreover,
 to fit empirical  hyperdegree distributions, we choose specific parameters make a large fraction of nodes only belong to  one  hyperedge.
Meanwhile, most modelled hyperedges contain one lead node, and only early lead nodes and a few nodes close to them can be persistently
contained by new hyperedges.
Those yield that the small/large degree nodes contribute positively/negatively  to degree assortativity and high clustering. Hence the model well reproduces  the transitions.
 In addition,  the tails of $C(k)$
   proportional to $1/k$ also holds in   modelled networks. For a lead node $i$, the probability of its new team member coauthor with the formers is
$P=\int \alpha \mu t^\beta_i/t^\beta   dt \propto t^\beta_i\propto 1/k_i$, where $k_i$ is the degree of node $i$.

\section*{Discussions and conclusions}

Five
typical mechanisms of cooperation evolution    hold for academic collaborations, which inspires
us to explore   game  mechanisms    in the   evolution of coauthorship networks.
We define a    cooperative game model on a circle, and
 reveal how the
costs and benefits  of   individuals generate a
 range of statistical and topological features of coauthorship networks, such as fat-tail, small-world, etc.
 It overcomes the  weakness of the   model in Ref.~\cite{Xie3}, a lot of parameters, and has the new ability to fit the
distribution  of paper team sizes and that of hyperdegree.
 Moreover, it
  has the
potential to illuminate specific   views and  implications in the broader study of cooperative  behaviors as follows.

Do there   exist innate rules behind the social complexity? It
provides an example of how
individual  strategies based on
maximizing benefit-minus-cost and on specific randomness
generate the   complexity emerged
in coauthorship networks. The general idea of the model
 potentially bridges
cooperative game theory and  specific social networks generated by human strategies,
e.~g. social affiliation networks.

 Does utilitarianism  help  the development of sciences?
The strategy of maximizing benefit-minus-cost will give rise to
flocking  to famous research team or to
hot fields. Taking such strategy helps to  collect publications and citations,
but      suppresses  diversity, and consequently does harm to the   flexibility of   academic  environment.
 However,   current academic evaluation methods and funding mechanisms
are mainly oriented by specific indexes, e. g. the number of citations.
 Specific regulations could be simulated through    the model to work out  a way
 to maintain   the balance in    academic environment, while
encouraging breakthroughs in key fields.

\section*{Acknowledgments}  This work is supported by    National   Science Foundation of China (Grant No. 61773020).

\section*{Authors' contributions}
The   authors have contributed equally to this work. All authors conceived and designed the research. ZX, JPL and ML wrote the paper. ZX analyzed the data.   All authors discussed the research and approved the final version of the manuscript.

\section*{Appendix}

\subsection*{Detecting boundary points.}

The following  boundary    detection  algorithms come from  Ref.~\cite{Xie6}.
\begin{table*}[!ht] \centering \caption{{\bf A boundary detection algorithm for probability density functions.} }
\begin{tabular}{l r r r r r r r r r} \hline
Input: Observations  $D_s$, $s=1,...,n$,  rescaling function $g(\cdot)$, and fitting model     $h(\cdot)$.\\
\hline
For   $k$ from $1$ to $\max(D_1,...,D_n)$ do: \\
~~~~Fit   $h(\cdot)$  to   the PDF $h_0(\cdot)$ of    $\{D_s, s=1,...,n|D_s \leq k\}$    by  maximum-likelihood\\ estimation; \\
~~~~Do    KS test for two    data
     $g(h(t))$ and $g( h_0(t))$, $t=1,...,k$ \\
   with the null hypothesis  they coming from the same continuous distribution;\\
~~~~Break  if  the test rejects the null hypothesis  at     significance level $5\%$. \\ \hline
Output: The current $k$ as the   boundary point. \\ \hline
 \end{tabular}
    \begin{flushleft}
    \end{flushleft}
\label{tab5}
\end{table*}

\begin{table*}[!ht] \centering \caption{{\bf Boundary point  detection algorithm for general functions.} }
\begin{tabular}{l r r r r r r r r r} \hline
Input: Data vector  $h_0(s)$, $s=1,...,K$, rescaling funtion $g(\cdot)$, and fitting model     $h(\cdot)$.\\
\hline
For   $k$ from $1$ to $K$ do: \\
~~~~Fit   $h(\cdot)$  to    $h_0(s)$, $s=1,...,k$  by regression; \\
~~~~Do  KS test for two    data vectors
     $g(h(s))$ and $g( h_0(s))$, $s=1,...,k$ with   the null\\
 hypothesis they coming from the same continuous distribution;\\
~~~~Break  if  the test rejects the null hypothesis  at   significance level  $5\%$. \\ \hline
Output: The current $k$ as the  boundary point. \\ \hline
 \end{tabular}
\label{tab6}
\end{table*}

\end{document}